\documentclass[a4paper,11pt]{article}

\usepackage{jcappub}
\usepackage{amsmath, amssymb}
\usepackage[sort&compress,numbers]{natbib}
\usepackage{txfonts}
\usepackage{siunitx}
\usepackage{graphicx}
\usepackage{mathtools}
\usepackage[utf8]{inputenc}

\usepackage{color}
\usepackage{epstopdf}

\allowdisplaybreaks

\title{Constraining the cosmological evolution of scalar-tensor theories with local measurements of the time variation of G}

\author[a]{Clare Burrage,}
\author[a,1]{Johannes Dombrowski,\note{Corresponding author.}}

\affiliation[a]{School of Physics and Astronomy, University of Nottingham,\\Nottingham, NG7 2RD, UK}

\emailAdd{Clare.Burrage@nottingham.ac.uk}
\emailAdd{Johannes.Dombrowski@nottingham.ac.uk}

\abstract{We determine the conditions for which the constraints from lunar laser ranging on the time evolution of the {\it local} gravitational constant can be extrapolated to impose  constraints on the time evolution of the {\it cosmological} gravitational constant in scalar-tensor theories of modified gravity. We allow for the possibility that the scalar-tensor theories are non-linear and contain a screening mechanism. This results in strong late Universe constraints on the running of the cosmological Planck mass described by the Horndeski function $|\alpha_M|\lesssim 0.002$. We find that our assumptions are valid for most Vainshtein and kinetic screening models, where the shift symmetry $\phi\rightarrow\phi + c$ holds, but are violated by some Chameleon and Symmetron screening models, where the macroscopic equivalence principle is broken.}


\arxivnumber{2004.14260}

\begin{document}
\maketitle
\flushbottom

\section{Introduction}

General relativity has, so far, stood the test of time. A plethora of high precision measurements in the Solar System, in laboratories, astrophysics, gravitational wave physics and cosmology \cite{Will:2014kxa, Adelberger:2009zz, Pulsars, Monitor:2017mdv, Ferreira:2019xrr} have shown excellent agreement between data and theory. On cosmological scales, the concordance model, which assumes general relativity with a cosmological constant, has been remarkably successful in describing the expansion history of the Universe and the evolution of cosmic structures \cite{Akrami:2018vks}, although recently some discrepancies, including the $H_0$ tension \cite{Verde:2019ivm} and the very mild $S_{\! 8}$ tension \cite{Abbott:2017wau} have been pointed out that may hint at physics beyond general relativity if systematics and other sources of error can be decisively excluded.

Despite these immense successes, general relativity eludes a unification with the other fundamental forces of physics and the inclusion of the cosmological constant in order to explain the late-time acceleration of the Universe requires an unnatural amount of fine tuning. Therefore, it is essential that we continue to test gravity and consider possible alternatives or extensions to general relativity \cite{Clifton:2011jh, Joyce:2014kja, Ishak:2018his}.

A particularly popular modification of general relativity is the addition of a `fifth force', which is mediated by a scalar field and which, similarly to gravity, acts between massive particles, see \cite{Joyce:2014kja} for a review. Typically, the scalar field $\phi$ is assumed to couple conformally to matter:
\begin{equation}
S=\int \mathrm{d}^4x\sqrt{-g}\left[ M_p^2\left( \frac{R}{2}-\Lambda_c \right) +\mathcal{L}_\phi \right] + S_{\! m}\left[ \Omega(\phi)g_{\mu\nu}, \psi_m \right], \label{general_action}
\end{equation}
where $\mathcal{L}_\phi$ is the scalar field Lagrangian and $S_{\! m}$ is the Lagrangian for the matter fields $\psi_m$, which are subject to the Jordan-frame metric $\Omega(\phi)g_{\mu\nu}$ ($g_{\mu\nu}$ is the Einstein-frame metric). $M_p^2=1/8\pi G_N$ is the reduced Planck mass squared defined in terms of Newton's gravitational constant $G_N$ and $\Lambda_c$ is an optional cosmological constant.

Since matter particles move on geodesics of the Jordan-frame metric, a non-relativistic test mass will, in addition to the gravitational acceleration, experience an acceleration due to the gradient of the conformal factor $\Omega(\phi)$:
\begin{equation}
\vec{a}_5=-\frac{1}{2}\vec{\nabla}\log\Omega. \label{acceleration}
\end{equation}
Since these fifth forces are heavily constrained in the Solar System, if they are to play a role cosmologically, they must be equipped with a screening mechanism, which weakens the fifth force on solar system scales. This can be achieved if non-linearities in the theory become important in specific regimes, for instance in high density regions like the Solar System, but are negligible on cosmological scales. The most common screening mechanisms are the Chameleon \cite{Khoury:2003rn}, Symmetron \cite{Hinterbichler:2010es}, Vainshtein \cite{Vainshtein1972393} and kinetic \cite{Babichev:2009ee} screening mechanisms, which we will discuss in more detail in Sections \ref{sec:Vainshtein} and \ref{sec:Chameleon}.

In a theory with conformal coupling, the observed gravitational constant will be modified by the conformal factor $\Omega$:
\begin{equation}
G(\phi)=\Omega(\phi) G_N. \label{grav_constant}
\end{equation}
This can be seen for example by transforming the action \eqref{general_action} into the Jordan frame, where the Einstein-Hilbert term (among other things) is multiplied by a factor of $\Omega^{-1}$. This gives rise to a variation of Newton's constant on both cosmological and solar system scales. 

 If the gravitational constant varies on  scales where cosmological perturbation theory remains linear, this variation is characterised by the  Horndeski parameter\footnote{The four Horndeski $\alpha$-functions ($\alpha_M$, $\alpha_B$, $\alpha_K$ and $\alpha_T$) were introduced in \cite{Bellini:2014fua} as independent functions describing the linear growth of structure in any Horndeski theory \cite{Horndeski:1974wa, Deffayet:2011gz} without redundancy. There is a translation between the Horndeski $\alpha$-functions and the effective field theory of dark energy parameters, see Ref.\ \cite{Frusciante:2019xia} for a review.}:
\begin{equation}
\alpha_M\coloneqq -H^{-1}\dot{G}/G\;.
\end{equation}
On cosmological scales, a time varying gravitational constant can provide an  explanation for the late-time accelerated expansion of the Universe alternative to the cosmological constant or the quintessence scenario, and is called self-acceleration, see e.g.\ \cite{Nicolis:2008in}. Here the expansion of the Universe is accelerated only in the (observed) Jordan frame, but not in the Einstein frame. This solution is characterised by the Horndeski parameter $\alpha_M$
being of order $1$. Recently, self-acceleration and modified gravity in general has come under heavy pressure from a combination of cosmological data including CMB, BAO and ISW data \cite{Renk:2017rzu}, the multi-messenger observation of a neutron star merger \cite{Baker:2017hug, Creminelli:2017sry, Sakstein:2017xjx, Ezquiaga:2017ekz} and theoretical constraints from gravitational wave instabilities \cite{Creminelli:2019kjy, Noller:2020afd}. For the most recent constraints combining these evidences see \cite{Noller:2020afd}, which essentially rules out $\alpha_M= \mathcal{O}(1)$. Predating this, Ref.\ \cite{khoury_nogo} argues that self-acceleration in Chameleon and Symmetron models is ruled out as soon as we require  that the fifth force is smaller than the gravitational force everywhere.

Within the Solar System the $\phi$ dependence of the gravitational constant in Eq.\ \eqref{grav_constant} is subject to constraints on the time evolution of the gravitational constant, in particular from lunar laser ranging \cite{lunarlaser_new}:
\begin{equation}
\left|\frac{\dot{\Omega}}{\Omega}\right|=\left|\frac{\dot{G}}{G}\right|\lesssim 0.002H_0. \label{lunar_laser}
\end{equation} 
This constraint on the time evolution of the conformal factor $\Omega$ is a priori only valid on solar system scales. The goal of this paper is to show under which circumstances these constraints from solar system scales can be used to constrain the cosmological evolution of the gravitational constant, i.e.\ $\alpha_M$, in the late Universe. Models only varying the gravitational constant in the early Universe, see e.g.\ Ref.\ \cite{Barrow:1997qh} for a review and Refs.\ \cite{Braglia:2020iik, Ballesteros:2020sik} for recent developments, are not affected by our constraints. Constraints on the variation of the gravitational constant in the early Universe come from analyses of the Big Bang Nucleosynthesis \cite{Alvey:2019ctk} and the CMB \cite{Ooba:2017gyn}.

The central idea of this paper, which is described in detail in Section \ref{sec:prove}, is very similar to the approach used in Ref.\ \cite{khoury_nogo}: If we assume that the acceleration of a test particle due to the fifth force, in Eq.\ \eqref{acceleration}, is always small compared to the acceleration due to the gravitational force $-\nabla\phi_N$, where $\phi_N$ is the gravitational potential, we can integrate both quantities from far outside the Milky Way to inside the Solar System. This leads to the statement that the local solar system value and the cosmological value of the conformal factor, $\Omega_l$ and $\Omega_0$ respectively, can only deviate by a term proportional to the local value of the gravitational potential, which is of order $10^{-6}$ in the Solar System:
\begin{eqnarray}
\left| \Omega_l - \Omega_0 \right|\lesssim 10^{-6}\Omega_0.
\end{eqnarray}
Therefore, the constraints from lunar laser ranging in Eq.\ \eqref{lunar_laser} on $\Omega_l$ also apply to the cosmological solution $\Omega_0$ and thus $\alpha_M$. While it was already shown in \cite{Babichev:2011iz, Kimura:2011dc, Barreira:2013xea, Brax:2018zvh} that theories invariant under the shift symmetry $\phi\rightarrow\phi+c$ are subject to the constraints from lunar laser ranging, our result serves as a confirmation of their results with an independent approach and applies to a wider range of fifth force models.

Our analysis depends on a number of assumptions, which we detail in Section \ref{sec:assumptions} and whose validity will depend on the specific fifth force model under consideration. While we find that our assumptions are valid for most theories which obey the shift symmetry $\phi\rightarrow\phi+c$ (this includes the most common Vainshtein and kinetic screening models), they are not exclusively valid for shift-symmetric theories. We carefully examine under which conditions non shift-symmetric theories like the Chameleon and Symmetron models could potentially escape our constraints due to a violation of the macroscopic equivalence principle, see Section \ref{sec:Chameleon}. With regard to the Chameleon and the Symmetron, our conclusions are thus more conservative than the claim in \cite{khoury_nogo}, which rules out self-acceleration entirely for these models.

This paper is structured as follows: Section \ref{sec:assumptions} summarises our key assumptions, Section \ref{sec:prove} proves that under those assumptions the lunar laser ranging constraints of Eq.\ \eqref{lunar_laser} imply strong constraints on the Horndeski parameter $\alpha_M$ and Section \ref{sec:discussing_assumptions} discusses the assumptions one by one in light of the most common fifth force models. We end with a conclusion in Section \ref{sec:conclusion}.

\section{From the Solar System to cosmological scales}

In this section we will prove, with minimal assumptions, that the deviation of the local conformal factor $\Omega_l$ from the cosmological average $\Omega_0$ must be small. Therefore constraints on the time evolution of the Planck mass from local solar system observations also constrain the cosmological evolution of the Planck mass for fifth force models respecting those assumptions. While our formal proof in Section \ref{sec:prove} will be completely model independent, it relies on a number of assumptions, which are summarized in Section \ref{sec:assumptions} and whose validity has to be checked for any fifth force model individually.

\subsection{Model dependent assumptions}
\label{sec:assumptions}

So far, observations from cosmological to local solar system scales have not detected a fifth force with high significance\footnote{The exception to this is Ref.\ \cite{Ferreira_galaxy}, which detects behaviour consistent with a weak screened fifth force, although  the effects could also be due to galactic physics. If these observations are confirmed, the force is still weak compared to gravity and so our analysis still applies.}, which makes it very unlikely that a fifth force, which is significantly stronger than gravity could be observed on any of these scales. The situation might be different on laboratory scales where strong fifth forces could still exist in Chameleon or Symmetron scenarios \cite{Burrage:2014oza}, although even here experimental constraints are strong \cite{Burrage:2017qrf}. However, these scales are not relevant for the purposes of this work. We therefore arrive at our first assumption: 
\begin{equation}
\textbf{Weak fifth force assumption: } \quad\left|\vec{F}_5\right| < \beta\left| \vec{F}_N \right|, \label{assumption1}
\end{equation}
which is assumed to be valid from cosmological ($\sim \SI{100}{Mpc}$)\footnote{The scale of the largest observed structures in the Universe \cite{Tully:2014gfa}.} to solar system scales ($\sim \SI{1}{AU}\approx \SI{5e-6}{pc}$). We define $\vec{F}_N$ as the Newtonian gravitational force and $\beta$ is a constant roughly of order $1$. We will actually see later (in the discussion around Eq.\ \eqref{second_option_constraint}) that $\beta$ could be as large as $10$ for our purposes.

Next, we assume that the weak fifth force assumption can be directly translated into a constraint on the acceleration of test particles:
\begin{equation}
\textbf{Equivalence principle assumption: } \quad \frac{1}{2}\left| \nabla\log \Omega \right| < \beta \left| \nabla\phi_N \right|, \label{assumption_2}
\end{equation}
with $\phi_N$ being the Newtonian potential. This step will in general not be valid for fifth force models which break the macroscopic equivalence principle. We will discuss this in great detail in Section \ref{sec:discussing_assumptions}. The validity of this assumption is therefore model dependent.

Furthermore, we assume that the fifth force and the gravitational force are parallel throughout the space time region we consider. This is motivated by the fact that both the fifth force and the weak field gravitational force are sourced in the same way by matter and that scalar mediated fifth forces are  attractive\footnote{Unless the equation of state parameter of the scalar field is phantom \cite{Amendola:2004qb}, in which case the theory is  unstable. Another possible exception is described in \cite{Wittner:2020yfc}.} \cite{Amendola:2017orw}. We can therefore write:
\begin{equation}
\textbf{Parallelism assumption: } \quad \frac{1}{2} \nabla\log \Omega = \beta(\vec{x})  \nabla\phi_N, \label{third_assumption}
\end{equation}
where we have promoted the constant $\beta$ to a function of space-time which fulfils:
\begin{equation}
0<\beta(\vec{x})\leq \beta_\text{max}\lesssim 10, \quad \forall \vec{x}.
\end{equation}

Finally, we assume that the time today, $t_0$, is not a special point in the evolution of the Universe. This is an important assumption because the fifth force on solar system scales can not be tested over cosmological time scales. More formally, we assume that Eq.\ \eqref{third_assumption} is valid for an extended period of time $\Delta t$ on cosmological scales (we will specify this more explicitly around Eq.\ \eqref{second_option_constraint}): 
\begin{equation}
\textbf{Naturalness assumption: } \quad \frac{1}{2} \nabla\log \Omega = \beta(\vec{x})  \nabla\phi_N  \quad \text{for}\quad t\in\left[ t_0-\Delta t:t_0 \right]). \label{final_assumption}
\end{equation}
Not only would it be unnatural for the strength of the fifth force to be suppressed only during a short (with respect to cosmological time scales) time interval around $t_0$, but we are also not aware of a realistic cosmological model predicting this.

We will make two additional very technical assumptions in the following, see Eqs.\ \eqref{integral_estimation} and \eqref{solar_system_estimate} and the discussions before these equations. These assumptions are however independent of the fifth force model, and therefore, are not mentioned here.

\subsection{Model independent proof}
\label{sec:prove}

We split the conformal factor $\Omega(\vec{x}, t)$ into its spatially averaged value $\Omega_0(t)$ plus an inhomogeneous part $\omega(\vec{x}, t)$:
\begin{equation}
\Omega=\Omega_0(t)+\omega(\vec{x}, t), \qquad \text{with}\quad \langle\omega(\vec{x}, t)\rangle=0, \label{decomposition}
\end{equation}
where $\langle\dots\rangle$ represents a spatial average. The function $\Omega_0(t)$ represents the cosmological background solution, and $\omega$ describes inhomogeneities sourced by the highly non-linear density distribution in the late Universe. 

We first show that on solar system scales $\omega$ has to be small compared to $\Omega_0$, meaning that the local conformal factor has to be very close to the cosmological conformal factor. For this we integrate the final assumption, Eq.\ \eqref{final_assumption}, along a path $\gamma$ from a point $\vec{x}_2$ far outside an overdensity, such as the Milky Way, to a point $\vec{x}_1$ inside the Solar System:
\begin{equation}
\frac{1}{2}\log \frac{\Omega(\vec{x}_1)}{\Omega(\vec{x}_2)}=\int_\gamma \beta(\vec{x})\nabla\phi_N\cdot\mathrm{d}\vec{x}. \label{integrating_the_potential}
\end{equation}
The integral on the right-hand side must be independent of the path $\gamma$ between the points $\vec{x}_2$ and $\vec{x}_1$ since $\beta(\vec{x})\nabla\phi_N$ has to be a conservative vector field. We now make the technical assumption that we can choose a path $\gamma$ where $\phi_N$ is decreasing from $\vec{x}_2$ to $\vec{x}_1$. This assumption will certainly hold for an isolated overdensity whose density increases monotonically towards the center of the object (such monotonically increasing density profiles are expected for any self-gravitating object). Since $\beta(\vec{x})>0$, the integral in Eq.\ \eqref{integrating_the_potential} has to be smaller than $0$ and we can estimate:
\begin{eqnarray}
0 >\int_\gamma \beta(\vec{x})\nabla\phi_N\cdot\mathrm{d}\vec{x} \geq \beta_\text{max} \int_\gamma \nabla\phi_N\cdot\mathrm{d}\vec{x} = \beta_\text{max} \left(\phi_N(\vec{x}_1)-\phi_N(\vec{x}_2)\right). \label{integral_estimation}
\end{eqnarray}
This statement can be made far more precise and constraining for a specific fifth force model, where $\beta(\vec{x})\ll \beta_\text{max}$ in screened regions. We demonstrate this for a simple cubic Galileon model in Appendix \ref{sec:galileon}. Therefore, we believe that Eq.\ \eqref{integral_estimation} should be considered a very conservative estimate.

We now choose the point $\vec{x}_2$, which so far we have only assumed to lie far outside the overdensity, such that $\Omega(\vec{x}_2)=\Omega_0$. Furthermore, we make the assumption that\footnote{This is the second and final technical assumption that we hinted at in Section \ref{sec:assumptions}.} $|\phi_N(\vec{x}_2)|\ll |\phi_N(\vec{x}_1)|$. Typical values for the gravitational potential inside of the Milky Way or the Solar System are of order $0>\phi_N(\vec{x}_1)\gtrsim\SI{-e-6}{}$. We therefore arrive at:
\begin{equation}
0>\frac{1}{2}\log \frac{\Omega(\vec{x}_1)}{\Omega(\vec{x}_2)}=\frac{1}{2}\log \left(1+\frac{\omega(\vec{x}_1)}{\Omega_0}\right) \gtrsim -\SI{e-6}{}\beta_\text{max}. \label{solar_system_estimate}
\end{equation}
Since the absolute value of the right-hand side is small compared to $1$, we can Taylor expand the logarithm and multiply by $2\Omega_0$:
\begin{equation}
0 > \omega_l \gtrsim -\SI{2e-6}{}\beta_\text{max}\Omega_0,\label{frame_equality}
\end{equation}
where $\omega_l\coloneqq\omega(\vec{x}_1)$ is a typical local (solar system scale) value of $\omega$. This proves that, even using the most conservative estimate in Eq.\ \eqref{integral_estimation}, $\omega$ on solar system scales has to be small compared to $\Omega_0$.

The constraints from lunar laser ranging, in Eq.\ \eqref{lunar_laser}, on the evolution of the local gravitational constant demand:
\begin{equation}
0.002 H_0\gtrsim \left|\frac{\dot{\Omega}(\vec{x}_1)}{\Omega(\vec{x_1})}\right|=\left|\frac{\dot{\Omega}_0+\dot{\omega}_l}{\Omega_0+\omega_l}\right|\approx \left|\frac{\dot{\Omega}_0+\dot{\omega}_l}{\Omega_0}\right|. \label{LunarLaser}
\end{equation}
These constraints are technically only valid over the time period during which lunar laser ranging tests were performed, i.e.\ over a few decades. However, in accordance with our naturalness assumption (see Eq.\ \eqref{final_assumption}), we assume that the constraints are valid for the extended cosmological time period $\Delta t$. 

There are now two possibilities for the two terms $\dot{\Omega}_0/\Omega_0$ and $\dot{\omega}_l/\Omega_0$ on the right-hand side of Eq.\ \eqref{LunarLaser}:
\begin{itemize}
	\item Both terms are individually smaller than the left-hand side of Eq.\ \eqref{LunarLaser}. In this case we conclude that the cosmological evolution of the gravitational constant, typically characterised by the Horndeski parameter $\alpha_M$, is constrained in the same way as the evolution of the local gravitational constant:
	\begin{equation}
	\left|\alpha_M\right|\coloneqq H^{-1}\left|\frac{\dot{\Omega}_0}{\Omega_0}\right|\lesssim 0.002.
	\end{equation}
	\item The two terms on the right-hand side of Eq.\ \eqref{LunarLaser} cancel each other, i.e.\ $\dot{\omega}_l\approx-\dot{\Omega}_0$. By assumption this cancellation has to be valid for an extended period of time $\Delta t$. Assuming that $\dot{\Omega}_0$ can be treated as approximately constant over a time interval $\delta t\leq\Delta t$, we can integrate $\dot{\omega}_l$:
	\begin{equation}
	\omega_l(t_0)-\omega_l(t_0-\delta t)\approx -\int_{t_0-\delta t}^{t_0}\dot{\Omega}_0(t_0)\mathrm{d}t\approx -\dot{\Omega}_0(t_0)\delta t. \label{constant_Omega_dot}
	\end{equation} 
	Since $\delta t<\Delta t$, both terms on the left-hand side have to fulfil Eq.\ \eqref{frame_equality} and we obtain:
	\begin{equation}
	\frac{\left| \dot{\Omega}_0(t_0) \right|}{\Omega_0}\lesssim \SI{2e-6}{}\frac{\beta_\text{max}}{\delta t}. \label{order_one_parameters}
	\end{equation}
	We can assume here that the approximation $\dot{\Omega}_0\approx \text{const}$ used in Eq.\ \eqref{constant_Omega_dot} is valid for at least a cosmologically short time interval $\delta t$ of order $0.01 H_0^{-1}$. Otherwise, the strong change in the evolution of $\Omega_0$ in the recent history of the Universe would make this small period of time a special point in the evolution of the Universe, which violates our Naturalness assumption. Therefore, the constraint
	\begin{equation}
	|\alpha_M|\lesssim 0.002 \label{second_option_constraint}
	\end{equation}
	still holds as long as $\beta_\text{max}\lesssim 10$. In other words the time interval $\Delta t$, during which the fifth force is assumed to be weak in our naturalness assumption \eqref{final_assumption}, has to fulfil $\Delta t>\delta t=0.001 H_0^{-1}\beta_\text{max}$ for our constraints on $\alpha_M$ to be valid.
\end{itemize}
We conclude that according to our assumptions, the evolution of the cosmological gravitational constant is strongly constrained:
\begin{equation}
|\alpha_M|\lesssim 0.002. \label{main_result}
\end{equation}
This is the central result of this paper. This bound on $\alpha_M$ is a significant improvement over previous bounds \cite{Noller:2020afd} and is independent of parametrisations of $\alpha_M$, but relies on the assumptions summarised in Section \ref{sec:assumptions}.

\section{Discussion of the central assumptions}
\label{sec:discussing_assumptions}

Although our proof in Section \ref{sec:prove} is independent of the underlying fifth force model, the assumptions summarized in Section \ref{sec:assumptions} are model dependent statements. Therefore, it is possible to evade the constraints from lunar laser ranging if the fifth force model violates one or more of these assumptions. In this section we will discuss these assumptions for theories of scalar-tensor modified gravity in general, and focus in particular on the most common screening models, Vainshtein, kinetic, Symmetron and Chameleon screening. For Vainshtein screening the fifth force is screened in regions where non-linearities in the second derivative of the scalar field become important, for kinetic screening non-linearities in the first derivatives are responsible for the screening and for Chameleon and Symmetron models screening becomes effective in regions of high matter density.

\subsection{Weak fifth force assumption}
\label{sec:assumption1}

The weak fifth force assumption in Eq.\ \eqref{assumption1} states that the fifth force between two objects whose size range from solar system to cosmological scales can not be significantly larger than the gravitational force between the two objects. We saw in our discussion around Eq.\ \eqref{order_one_parameters} that `significantly larger' means more than an order of magnitude larger than the gravitational force. The goal of this section is to summarise some of the observational bounds on the strength of a fifth force.

On solar system scales a strong fifth force is decisively ruled out by precision measurements of general relativity in the Solar System \cite{Will:2014kxa}. On galactic scales some evidence has been found for a very weak screened fifth force \cite{Ferreira_galaxy}, but according to the authors it is possible that a better understanding of galaxy formation would weaken the evidence. Even if a weak force of this form exists, it still satisfies our assumption that a fifth force, if it exists, can't be substantially stronger than gravity. In general, it is difficult to make definite statements about the presence of a fifth force on galactic scales because of degeneracies with uncertainties in our knowledge of the physics of galaxy formation. However, it seems unlikely that a fifth force that is an order of magnitude stronger than the gravitational force would have evaded detection.

For a fifth force model which screens effectively on small scales, such as within the Solar System, the strongest constraints on the strength of a fifth force come from observations of the largest scales in the universe, where the growth of structure can be treated linearly. On these scales screening is generally less effective than on smaller scales due to the low densities of the density perturbations on these scales. Therefore, we expect the effects of the fifth force to be strongest there. A strong fifth force would significantly modify the linear growth of structure $D_+(a)$, which is described by the differential equation:
\begin{equation}
D_+^{\prime\prime}+\mathcal{H}D_+^\prime = \frac{3}{2}\mathcal{H}^2\Omega_m(a)\left(1+\beta(a)\right)D_+, \label{growth_function}
\end{equation}
where primes denote derivatives with respect to conformal time, $\mathcal{H}$ is the conformal Hubble function, $\Omega_m(a)$ is the fractional matter density and $\beta(a)$ is the relative strength of the fifth force with respect to the gravitational force on linear scales as a function of the scale factor $a$. The function $\beta(a)$ is strongly constrained by observations of redshift space distortions. The study in Ref.\ \cite{DES_year1} using data from the DES 1-year results constrains $\beta(a)$ (called $\mu(a)$ in Ref.\ \cite{DES_year1}) and making use of the parametrisation $\mu(a)=\mu_0\Omega_\Lambda(a)/\Omega_{\Lambda,0}$, concludes: $\mu_0=-0.11^{+0.42}_{-0.46}$. Similarly, Ref.\ \cite{Mueller:2016kpu} constrains $G_M\coloneqq1+\beta(a)$ by binning it into two redshift intervals and obtains $G_M(z<0.5)=1.26\pm0.32$ and $G_M(z>0.5)=0.986\pm0.022$. This clearly rules out fifth forces with  strength relative to gravity of $\beta\sim 10$.

It is important to make one caveat when using redshift space distortion data in order to constrain modified theories of gravity. Several steps in the analysis of redshift space distortion data assume a certain cosmology, typically $\Lambda$CDM. The effects of this assumption have been estimated in \cite{Barreira:2016ovx, Bose:2017myh} and were shown to be important for future galaxy surveys, but are negligible for the precision of current data. Therefore, this should not affect the validity of the weak fifth force assumption.

Finally, we note that it is very unlikely that a theory of modified gravity could have a fifth force which is stronger than gravity on linear scales without significantly changing the expansion history of the Universe, which is well constrained. This makes a violation of the weak fifth force assumption even more unlikely.

\subsection{Equivalence principle assumption} 
\label{sec:equivalence_principle}

The Equivalence principle assumption of Eq.\ \eqref{assumption_2} is the assumption which is the most likely to be violated by theories of screened fifth forces. This assumes that small fifth forces imply small gradients of the conformal factor, a result which would follow immediately from the weak fifth force assumption \eqref{assumption1} if a macroscopic equivalence principle holds for all the astrophysical and cosmological objects used as tracers in tests of gravity. This is true for some, but not all, fifth force models as was shown in \cite{Equivalence_principle}. We will review the results of \cite{Equivalence_principle} here and discuss their implications on the validity of our constraints in Eq.\ \eqref{main_result} for different fifth force models depending on their screening models.

The fifth force on an extended, non-relativistic test object B due to another object A can be computed through:
\begin{equation}
F_5^i= -\int_S T_\phi^{ji}n_j\mathrm{d}S, \label{fifth_force_extended}
\end{equation}
where $S$ is a surface enclosing object B, $n$ is the unit vector normal to $S$ and $T_\phi$ is the energy-momentum tensor of the scalar field. We will assume spherical symmetry for the test object B in the following.

In simple scenarios we can solve the surface integral in Eq.\  \eqref{fifth_force_extended} analytically. Let us assume that the total gradient of the scalar field on the surface $S$ is well approximated by adding the gradients of the fields $\phi_A$ and $\phi_B$, where $\phi_A$ and $\phi_B$ are the field profiles we would compute around objects A and B if they were isolated. Furthermore, we assume that the field profile $\phi_B$ is well described by a $1/r$ power law and that the gradient of $\phi_A$ is constant over the surface $S$. With these assumptions the gradient of the scalar field on the surface $S$ is given by:
\begin{equation}
\partial_i\phi =\partial_i\phi_A + \partial_i\phi_B=\partial_i\phi_A +\frac{Q_B}{4\pi}\frac{x_i}{r^3}, \label{gradient_sum}
\end{equation}
where $Q_B$ is the scalar charge of object B. Finally, we assume that the energy-momentum tensor of the scalar field is given by:
\begin{equation}
T_\phi^{ij}=\partial^i\phi\partial^j\phi-\frac{1}{2}\delta^{ij}\partial_k\phi\partial^k\phi, \label{energy_momentum}
\end{equation}
where we have neglected time derivatives of the scalar field compared to spatial derivatives, which is reasonable for non-relativistic objects. Under these assumptions we find the simple result:
\begin{equation}
F_5^i= -Q_B\partial^i\phi_A. \label{fifth_force}
\end{equation}
We will use this result in the following two subsections in order to discuss the equivalence principle assumption in light of the most common fifth force models, which we characterise by their screening mechanisms.

\subsubsection{Equivalence principle assumption -- Vainshtein}
\label{sec:Vainshtein}

The first type of screening mechanisms we would like to discuss are Vainshtein \cite{Vainshtein1972393} and kinetic screening \cite{Babichev:2009ee}. For brevity we will refer to both of these screening mechanisms as `Vainshtein-type screening' in the following. These screening mechanisms weaken the fifth force in regions of high derivatives of the scalar field; second derivatives in case of Vainshtein screening and first derivatives for kinetic screening. Around matter sources, there will typically be a non-linearity radius $r_V$ (also Vainshtein radius), within which the fifth force is screened and outside which the fifth force behaves just like non-relativistic gravity. 

The action for Vainshtein-type screening models will typically be invariant under the shift symmetry $\phi\rightarrow \phi+c$ except for the conformal factor. Thus, the equation of motion can be written in terms of the current $J^\mu$:
\begin{equation}
\nabla_\mu J^\mu= \frac{\mathrm{d}\log \Omega^{1/2}}{\mathrm{d}\phi}\rho_m=\frac{\xi}{M_p}\rho_m, \label{eom_shift_symmetry}
\end{equation}
where, in the last equality, we made the common choice $\Omega^{1/2}(\phi)=\exp(\xi\phi/M_p)$ for the conformal factor such that the equation of motion for the scalar field is completely shift symmetric. We can integrate Eq.\ \eqref{eom_shift_symmetry} around a concentrated matter source of mass $M_B$ and obtain:
\begin{equation}
\int_S\vec{J}\cdot\vec{n}\,\mathrm{d}S = \frac{\xi}{M_p}M_B,
\end{equation}
assuming that the matter source is static. At distances larger than the Vainshtein radius $r_V$, the conserved current $J^\mu=\partial^\mu\phi$ and we obtain the far field:
\begin{equation}
\partial_r\phi_{B}(r>r_V)=\frac{\xi M_B}{4\pi M_p r^2} = \frac{Q_B}{4\pi}\frac{1}{r^2},
\end{equation}
where we defined the scalar charge $Q_B=\xi M_B/M_p$. Furthermore, in the regime $r>r_V$ the energy-momentum tensor is well approximated by Eq.\ \eqref{energy_momentum}. If we choose the surface $S$ for the integral in Eq.\ \eqref{fifth_force_extended} to lie outside of the Vainshtein radius, we can use Eq.\ \eqref{fifth_force} to calculate the fifth force acting on object B due to another object A if $\partial_i\phi_A$ can be treated as constant over the surface $S$, or in other words if the wavelength of the field $\phi_A$ is larger than the Vainshtein radius of object B. We also use that the symmetry $\partial_\mu\phi\rightarrow\partial_\mu\phi+c_\mu$ holds approximately for any of the theories under consideration for $r>r_V$. This enables us to add the gradient of $\phi_A$ to the gradient of $\phi_B$ as in Eq.\ \eqref{gradient_sum}. The validity of Eq.\ \eqref{fifth_force} is thus not restricted to theories which are completely invariant under the Galilean symmetry, but holds for any theory where the Galilean symmetry is approximately valid for $r>r_V$, which includes theories with kinetic screening.

Therefore, if the Vainshtein radius of object B is much smaller than the wavelength of the field $\phi_A$, the magnitude of the fifth force acting on object B relative to the gravitational force on B can be expressed through the gradient of the conformal factor $\Omega(\phi_A)$ at the position of object B:
\begin{equation}
\frac{\left|\vec{F}_5\right|}{\left|\vec{F}_N\right|}=\frac{\left|-Q_B\vec{\nabla}\phi_A\right|}{\left|-M_B\vec{\nabla}\phi_{N,A}\right|}=\frac{1}{M_p}\frac{\left|\xi\vec{\nabla}\phi_A\right|}{\left|\vec{\nabla}\phi_{N,A}\right|}=\frac{\left|\vec{\nabla}\log\Omega^{1/2}(\phi_A)\right|}{\left|\vec{\nabla}\phi_{N,A}\right|}, \label{equivalence_principle_derivative_screening}
\end{equation}
where $\phi_{N,A}$ is the gravitational potential of object A at the position of object B. We conclude that the equivalence principle assumption, Eq.\ \eqref{assumption_2}, directly follows from the weak fifth force assumption, Eq.\ \eqref{assumption1}, if the Vainshtein radius of the tracer object can be neglected.

Due to the nature of Vainshtein-type screening, the ratio $\nabla\Omega^{1/2}/\nabla\phi_N$ is largest at large distances from matter sources and for objects of low density\footnote{The Vainshtein radius scales with the total mass of the object, see Eq.\ \eqref{Vainshtein_radius}. For a low density object, the ratio between the Vainshtein radius and the extent of the object is therefore lower than for an object with high density.}. Therefore, if we constrain the ratio $\nabla\Omega^{1/2}/\nabla\phi_N$ through observations of the largest structures in the Universe, which are low in density, we have constrained the ratio $\nabla\Omega^{1/2}/\nabla\phi_N$ everywhere. For the largest structures, the Vainshtein radius of tracer galaxies can be neglected in most cases and thus Eq.\ \eqref{equivalence_principle_derivative_screening} will typically apply.\footnote{This picture was confirmed with simulations of large scale structures in \cite{Falck:2014jwa}.} We have seen in Section \ref{sec:assumption1} that the fifth force from the largest structures is well constrained by observations of redshift-space distortions.

We summarise that the equivalence principle assumption, Eq.\ \eqref{assumption_2}, holds for all Vainshtein-type screening models with only one exception: Depending on the parameters of the theory under consideration, it would in principle be possible for the Vainshtein radius of the tracer galaxies to be of the same scale as the largest structures in the Universe. In this case, the tracer might self-screen the fifth force with its own non-linear field \cite{Hiramatsu:2012xj} and Eq.\ \eqref{equivalence_principle_derivative_screening} would not be valid anymore. However, if the Vainshtein radius of single galaxies was as large as the largest structures in the Universe, the fifth force would be screened everywhere in the Universe. We are not aware of a study which analyses these highly non-linear theories and their impact on cosmic structure formation in detail. Thus, we draw the conservative conclusion that these theories, if they turn out to be consistent, might potentially escape our constraints.

\subsubsection{Equivalence principle assumption -- Chameleon}
\label{sec:Chameleon}

The second category of screening mechanisms, including the Symmetron and the Chameleon, will be called `Chameleon-type' screening in the following. Here the minimum of the effective potential of the scalar field $V^{\mathrm{eff}}(\phi)=V(\phi)+\Omega^{1/2}(\phi)\rho$, which depends on the local matter density, fixes the local value of the scalar field $\phi=\phi(\rho)$. In regions of high matter density the absolute value of the scalar field is small, which, for the Chameleon, leads to a high effective mass of the scalar field $m_\mathrm{eff}=V^\mathrm{eff}_{,\phi\phi}$ and therefore a short range of the fifth force, and for the Symmetron, gives a small coupling between matter and the scalar field. The conformal factor for the Chameleon is typically assumed to be: $\Omega^{1/2}(\phi)=\exp(\xi\phi/M_p)$; for the Symmetron however, the conformal factor is a function of $\phi^2$, i.e.\ $\Omega^{1/2}(\phi)=\exp(\xi^2\phi^2/2M_p^2)$.

To understand the field profile around compact objects in these models we follow the treatment in Refs.\ \cite{Joyce:2014kja, Khoury:2003rn}. We consider a static, spherically symmmetric matter source B with radius $R_B$ and constant density $\rho_B$ which is embedded in a homogeneous background density $\rho_{bg}$. We assume that the matter source is large or dense enough to be screened. For these screening mechanisms this means that the field value inside the object is given by the minimum of the effective potential $\phi(\rho_B)$. Far away from the object the field becomes $\phi_{bg}=\phi(\rho_{bg})$ which is typically much larger than the field inside the object $\phi(\rho_B)$. The solution that matches these two boundary conditions is approximately given by:
\begin{equation}
\phi_{B}(r>R_B)=\phi_{bg}-\frac{R_B}{r}(\phi_{bg}-\phi(\rho_B))\mathrm{e}^{-m(\rho_{bg})(r-R_B)}, \label{Chameleon_field_profile}
\end{equation}
assuming that the effective potential is well approximated by a quadratic outside the overdensity. We now write this solution in a more indicative way:
\begin{equation}
\phi_{B}(r>R_B)=\phi_{bg}-\lambda_B\frac{\xi M_B}{4\pi M_p}\frac{\mathrm{e}^{-m(\rho_{bg})(r-R_B)}}{r}, \label{chameleon_solution}
\end{equation}
where the screening factor $\lambda_B$ is defined as:
\begin{equation}
\lambda_B=\frac{3M_p(\phi_{bg}-\phi(\rho_B))}{\xi\rho_B R_B^2}=\frac{(\phi_{bg}-\phi(\rho_B))}{2\xi M_p\phi_N(R_B)}\approx\frac{\phi_{bg}}{2\xi M_p\phi_N(R_B)}. \label{lambda}
\end{equation}
We used the gravitational potential at the surface of the object $\phi_N(R_B)=M_B/8\pi M_p^2R_B$. It can be shown \cite{Burrage:2014oza} that the solution in Eq.\ \eqref{chameleon_solution} is consistent with our assumption that the field inside the object is given by the minimum of the effective potential $\phi(\rho_B)$ if the screening factor satisfies $\lambda_B\ll 1$. A small screening factor also means that the scalar charge $Q_B=\xi\lambda_BM_B/M_p$ is proportional to only a fraction of the entire mass of object B. Ref.\ \cite{Burrage:2014oza} also demonstrates that if the screening factor $\lambda_B$ computed through Eq.\ \eqref{lambda} would be larger than $1$, it has to be replaced by $1$. In this case we speak of an unscreened object.

We can now compute the fifth force acting on object B due to another object A by means of Eq.\ \eqref{fifth_force_extended} by choosing the surface $S$ just outside the object B such that the exponential decay in the solution \eqref{chameleon_solution} can be neglected:
\begin{equation}
F_5^i=-\lambda_BM_B\frac{\xi}{M_p}\partial^i\phi_A. \label{self_screening}
\end{equation}
The strength of the fifth force relative to the gravitational force acting on object B is thus given by:
\begin{equation}
\frac{\left|\vec{F}_5\right|}{\left|\vec{F}_N\right|}=\lambda_B\frac{\xi}{M_p}\frac{\left| \vec{\nabla}\phi_A \right|}{\left| \vec{\nabla}\phi_{N,A} \right|}, \label{relative_strength_general}
\end{equation}
where $\phi_A$ and $\phi_{N,A}$ are taken at the position of object B. Analogous to the Vainshtein case, see Eq.\ \eqref{equivalence_principle_derivative_screening}, we can express the right-hand side of Eq.\ \eqref{relative_strength_general} in terms of the derivative of the conformal factor $\Omega$. For the chameleon ($\log\Omega^{1/2}=\xi\phi/M_p$) we have:
\begin{equation}
\frac{\left|\vec{F}_5\right|}{\left|\vec{F}_N\right|}= \lambda_B\frac{\left|\vec{\nabla}\log\Omega^{1/2}(\phi_A)\right|}{\left|\vec{\nabla}\phi_{N, A}\right|}, \label{equivalence_chameleon}
\end{equation}
and for the Symmetron ($\log\Omega^{1/2}=\xi^2\phi^2/2M_p^2$):
\begin{equation}
\frac{\left|\vec{F}_5\right|}{\left|\vec{F}_N\right|}=\lambda_B\frac{M_p}{\xi\phi_A}\frac{\left|\vec{\nabla}\log\Omega^{1/2}(\phi_A)\right|}{\left|\vec{\nabla}\phi_{N, A}\right|}. \label{equivalence_symmetron}
\end{equation}
In contrast to the Vainshtein case a small fifth force, i.e.\ a small left-hand side in Eqs.\ \eqref{equivalence_chameleon} and \eqref{equivalence_symmetron}, does not necessarily mean that the gradient of the conformal factor has to be small. For the Chameleon it depends on the properties of the test object B whether the gradient of the conformal factor is constrained by observations of small fifth forces. If the test object is screened, $\lambda_B\ll 1$, the conformal factor is less constrained than if the test object is unscreened, $\lambda_B=1$. For the Symmetron it additionally depends on the field $\phi_A$ at the position of the test object.

Therefore, the equivalence principle assumption, Eq.\ \eqref{assumption_2}, can be violated by Chameleon-type screening, if all relevant astrophysical and cosmological test objects are screened. In this case the constraint on $\alpha_M$ in Eq.\ \eqref{main_result} could be violated by Chameleon-type screening models. This can be shown by integrating Eq.\ \eqref{relative_strength_general} in the same way we have integrated the equivalence principle assumption in Section \ref{sec:prove}: We define $\beta(\vec{x})\coloneqq\left|\vec{F}_5\right|/\left|\vec{F}_N\right|$, multiply by $\left|\vec{\nabla}\phi_{N,A}\right|$ and integrate from a point far outside of object A, where $\phi_A$ is equal to the cosmological solution $\phi_{bg}$, to a point inside object A, where $\phi_A$ is given by the local value of the scalar field $\phi_{l}$. Estimating the integral in the same way as in Section \ref{sec:prove} through the maximum value $\beta_\text{max}$ of $\beta(\vec{x})$, gives the following constraint on the difference between the cosmological and the local field:
\begin{equation}
0>\lambda_B\frac{\xi}{M_p}\left( \phi_{l} - \phi_{bg} \right)\gtrsim \beta_\text{max} \phi_{N,l}\sim -10^{-6}\beta_\text{max}, \label{frame_equality_chameleon}
\end{equation}
where we introduced the local gravitational potential $\phi_{N,l}$. For simplicity of the calculation we have assumed that the screening factor $\lambda_B$ is the same for all tracer objects which are used to test the fifth force on all relevant scales from far outside to inside object A.\footnote{This is a strong simplification and is not true in general, but it serves here to demonstrate the central idea.} If we could treat all tracer objects as unscreened, i.e.\ $\lambda_B=1$, Eq.\ \eqref{frame_equality_chameleon} becomes equivalent to Eq.\ \eqref{solar_system_estimate} in the sense that the value of the local conformal factor would again be constrained to be very close to the cosmological value for both the Chameleon ($\log\Omega^{1/2}=\xi\phi/M_p$) and the Symmetron ($(\log\Omega^{1/2})^{1/2}=\xi\phi/\sqrt{2}M_p$). In this case our constraints on $\alpha_M$ in Eq.\ \eqref{main_result} still hold. However, if the tracer objects are screened, i.e.\ $\lambda_B\ll 1$, the local and cosmological field values could potentially deviate enough to allow for a self-accelerating solution of the cosmological field characterised by $\alpha_M\sim 1$.

It has been pointed out in the literature, see e.g.\ Ref.\ \cite{Joyce:2014kja}, that $\lambda_B\ll 1$ is only possible for an extended period of time if the field excursion of the cosmological background field is small because $\lambda_B$ is proportional to $\phi_{bg}$, see Eq.\ \eqref{lambda}. This argument would suggest that $\alpha_M\sim 1$ is still ruled out even if the theory avoids the equivalence principle assumption and therefore our constraints in Eq.\ \eqref{main_result}. However, if $\xi\gg\phi^{-1}_N(R_B)$, $\lambda_B$ could remain small even if the field excursion is large, see Eq.\ \eqref{lambda}. Therefore, $\alpha_M$ is unconstrained for models with large coupling $\xi\gg 10^{6}$ since all tracers can be consistently screened for an extended period of time thereby violating our equivalence principle assumption.

In the related study Ref.\ \cite{khoury_nogo}, which establishes a no-go theorem for self-acceleration from Chameleon and Symmetron fields, the violation of the macroscopic equivalence principle was ignored with the argument that a huge backreaction effect on the expansion history is expected if all tracers were screened. While we agree that the background evolution should be reconsidered in this case, we are not aware of an argument showing that this reconsidered background evolution can not have $|\alpha_M|\sim 1$. Therefore, we would like to draw a more conservative conclusion by stating that a large $\xi$ could potentially invalidate our constraint in Eq.\ \eqref{main_result}. Hopefully, a future analysis of this backreaction effect will shine some light on this issue.

\subsection{Parallelism assumption}

The assumption that the fifth force is parallel to the gravitational force, Eq.\ \eqref{third_assumption}, is a technical assumption enabling us to make the simple estimate in Eq.\ \eqref{integral_estimation}. Small violations of this assumption are not problematic as long as the estimate \eqref{integral_estimation} still holds.

Since both the fifth force and the gravitational force are sourced by matter, they will always be parallel around spherically symmetric matter sources. The same is going to be true far away from a matter source, where the monopole of the matter distribution dominates. However, close to irregular matter distributions the two forces are only guaranteed to be parallel if they obey the same force law, i.e.\ the Poisson equation\footnote{We are assuming non-relativistic matter sources here because all objects of interest for our purposes, i.e.\ the Solar System, the Milky Way and other sub-horizon structures, are non-relativistic.} $\Delta\phi\propto\rho$.

For the Vainshtein-type screening mechanisms the Poisson equation will be approximately valid outside the Vainshtein radius, where the force is unscreened, and for Chameleon type screening the Poisson equation is a good approximate description of the field profile close to the matter source, where the Yukawa damping can be neglected.  Violations of the parallelism assumption, coming from deviations of the equation of motion from the Poisson equation, therefore occur only in regions where the gradient of the scalar field is suppressed compared to the gradient of the gravitational potential. Thus, the effect of these violations of the parallelism assumption should be irrelevant for the estimate in Eq.\ \eqref{integral_estimation}. For Vainshtein screening we show this explicitly in Appendix \ref{sec:galileon}, where Eq.\ \eqref{galileon_estimate} is the analogue of the estimate in Eq.\ \eqref{integral_estimation} and depends only on the fields outside the Vainshtein radius, i.e.\ where the scalar field and the gravitational field both obey a Poisson equation.

\section{Summary and Conclusions}
\label{sec:conclusion}

We have considered a general fifth force model, where a scalar field couples conformally to matter, and demonstrated that the constraints from lunar laser ranging on the time evolution of the local gravitational constant strongly constrain the evolution of the cosmological gravitational constant, i.e.\ $\alpha_M$, under a specific set of assumptions. We have assumed that 1.\ the fifth force is weak compared to the gravitational force on any scale from cosmological to solar system scales, see Eq.\ \eqref{assumption1}, that 2.\ a macroscopic equivalence principle holds for the objects used as tracers in tests of gravity on those scales, see Eq.\ \eqref{assumption_2}, that 3.\ the fifth force and the gravitational force are mostly parallel, see Eq.\ \eqref{third_assumption}, and that 4.\ all of those assumptions hold for an extended period of time on cosmological scales, see Eq.\ \eqref{final_assumption}. 

Furthermore, we made two model independent, technical assumptions in Section \ref{sec:prove}: If $\vec{x}_1$ is a point inside an overdensity like the Solar System and $\vec{x}_2$ is a point far outside the overdensity such that the conformal factor $\Omega(\vec{x}_2)$ is given by the cosmological average $\Omega_0$, we assumed that there exists a path $\gamma$ from $\vec{x}_1$ to $\vec{x}_2$ where the gravitational potential is monotonically increasing, and we assumed that $|\phi_N(\vec{x}_2)|\ll |\phi_N(\vec{x}_1)|$.

Under all of these assumptions we showed that the conformal factor in the Solar System has to be relatively close to the cosmological average, see Eq.\ \eqref{frame_equality}. Due to the strong constraints on the time evolution of the gravitational constant in the Solar System, the running of the Planck mass on cosmological scales in the late Universe is therefore heavily constrained: $|\alpha_M|\lesssim 0.002$. This is a significant improvement over previous constraints on $\alpha_M$ in the literature \cite{Noller:2020afd}, and furthermore has the advantage of being independent of a parametrisation of $\alpha_M$. Using this bound on the evolution of the cosmological gravitational constant should lead to significant improvements of constraints on cosmological models as is shown in, for example, Ref.\ \cite{Perenon:2019dpc}.

The validity of our assumptions should be considered for every fifth force model individually. For models of current interest, the most likely assumption to be invalid is the equivalence principle assumption. For shift-symmetric theories like Vainshtein and kinetic screening models, we argued that the macroscopic equivalence principle is valid for large scale structure tests of fifth forces, such as redshift space distortions, if the Vainshtein radius of the tracer galaxies can be assumed to be small compared to the large scale structures. For theories which predict the Vainshtein radius of these galaxies to be of the order of the largest structures in the universe, the macroscopic equivalence principle would break down and our constraints might not be valid. However, it remains to be seen if such a theory, which is non-linear everywhere, can be consistent with cosmological data. We therefore conclude that most Vainshtein and kinetic screening models should be subject to our constraints on $\alpha_M$. This validates the conclusions reached in \cite{Babichev:2011iz} with an independent approach and extends upon them since our analysis is also valid for some Chameleon-type screening mechanisms. 

Chameleon and Symmetron mechanisms can violate the macroscopic equivalence principle and therefore can, in some regions of their parameter space, evade our constraints if the coupling scale $\xi$ is large compared to the inverse gravitational potential on the surface of the tracer object. It remains an open question whether there might be a large backreaction effect for theories which violate the macroscopic equivalence principle as is suggested in \cite{khoury_nogo}.

We note that the gravitational constant determining the propagation of gravitational waves can in principle differ from the gravitational constant describing the strength of the gravitational force between massive objects, see Ref.\ \cite{Wolf:2019hun} where it is argued that this could lead to a novel way of probing the fifth force. The difference between these two couplings was calculated explicitly for the chameleon in Ref.\ \cite{Lagos:2020mzy}. The analysis in Ref.\ \cite{Lagos:2020mzy} uses that the local and cosmological conformal factors are close to each other, which we confirmed here for a small coupling $\xi$.

We close by remarking that modifications of gravity without a conformal coupling, for example theories with kinetic braiding \cite{Deffayet:2010qz}, where the gravitational force is modified through a mixing of the kinetic terms of the metric and the scalar field, are not affected by our constraints.

\section*{Acknowledgements}

We would like to thank Macarena Lagos for inspiring discussions in the early stages of this work. Furthermore, we would like to thank Ben Elder, Daniela Saadeh and Peter Millington for helpful discussions on the validity of our assumptions in different fifth force models. This work was supported by a Research Leadership Award from the Leverhulme Trust. CB is also supported by a Royal Society University Research Fellowship.

\appendix

\section{Cubic Galileon}
\label{sec:galileon}

We will show here how the conservative estimate, Eq.\ \eqref{integral_estimation}, can be made far more constraining for a cubic Galileon model. We define our model through the action \eqref{general_action} with the scalar Lagrangian:
\begin{equation}
\mathcal{L}_\phi= -\frac{1}{2}(\nabla\phi)^2 -\frac{1}{2M^3}\Box\phi(\nabla\phi)^2,
\end{equation}
where $M$ is a mass scale. We assume the conformal factor:
\begin{equation}
\Omega(\phi)=\exp \left( 2\xi\frac{\phi}{M_p} \right).
\end{equation}
With this conformal factor the scalar field equation of motion is shift-symmetric:
\begin{equation}
\Box\phi+\frac{1}{M^3}\left( \left(\Box\phi\right)^2-R_{\mu\nu}\nabla^\mu\phi\nabla^\nu\phi-\left( \nabla_\mu\nabla_\nu\phi \right)\left( \nabla^\mu\nabla^\nu\phi \right) \right)= \frac{\xi}{M_p}\rho, \label{field_eom}
\end{equation}
where $\rho$ is the non-relativistic matter density and $R_{\mu\nu}$ the Ricci tensor. 

For brevity, we consider a simple setting with an isolated, spherically symmetric overdensity $\rho_l(r)$ of radius $r_0$ on top the cosmological background density $\bar{\rho}(\tau)$, where $\tau$ is the conformal time. For the metric, we assume a FLRW metic with a small gravitational potential ($|\phi_N|\ll 1$):
\begin{equation}
\mathrm{d}s^2= a^2\left[ -(1+2\phi_N)\mathrm{d}\tau^2+(1-2\phi_N)\mathrm{d}\vec{x}^{\,2} \right].
\end{equation}
On the boundary ($r\rightarrow \infty$), the solution for $\phi$ has to approach the cosmological solution $\bar{\phi}$, which fulfils the background equation of motion:
\begin{equation}
\frac{\xi}{M_p}\bar{\rho}(\tau) = - \frac{1}{a^4}\frac{\partial}{\partial\tau}\left( a^2\bar{\phi}' \right) + \frac{3}{M^3a^4}\frac{\partial}{\partial\tau}\left( \mathcal{H}\bar{\phi}^{\prime 2} \right). \label{eom_background}
\end{equation}
We now make the ansatz:
\begin{equation}
\phi(r, \tau) = \bar{\phi}(\tau)+\phi_{l}(r, \tau) \qquad \text{with} \qquad \phi_l(r\rightarrow\infty, \tau) \rightarrow 0, \label{decomposition2}
\end{equation}
and make the quasi-static approximation, which states that time derivatives of the local solution $\phi_l$ can be neglected compared to spatial derivatives. Assuming a small gravitational potential $|\phi_N|\ll 1$, the equation of motion for the scalar field and the (0,0) component of the Einstein equations become:
\begin{eqnarray}
a^2r^2\frac{\xi}{M_p}\rho_l(r) &=& \frac{\partial}{\partial r} \left[\left(1-\frac{2}{M^3a^3}\frac{\partial}{\partial\tau}\left( a\bar{\phi}' \right)\right) r^2\phi_{l,r} +\frac{2}{a^2M^3} r\phi_{l,r}^2 -\frac{\bar{\phi}^{\prime 2}}{M^3a^2} r^2\phi_{N,r} \right],  \\
a^2r^2\rho_l(r)&=& \frac{\partial}{\partial r}\left[ 2M_p^2r^2\phi_{N,r} -\frac{\bar{\phi}^{\prime 2}}{M^3a^2} r^2 \phi_{l,r} \right]. \label{Poisson}
\end{eqnarray}
Combining the two equations and eliminating the gravitational potential yields:
\begin{equation}
\frac{\tilde{\xi}}{M_p}\rho_l(r) = \frac{\lambda(\tau)}{a^2r^2}\frac{\partial}{\partial r}\left( r^2\phi_{l,r} \right) +\frac{1}{a^4M^3}\frac{2}{r^2}\frac{\partial}{\partial r}\left( r\phi_{l,r}^2 \right), \label{master_equation}
\end{equation}
where we defined the useful quantities:
\begin{eqnarray}
\lambda(\tau)&\coloneqq& 1-\frac{2}{M^3a^3}\frac{\partial}{\partial\tau}\left( a\bar{\phi}' \right) -\frac{\bar{\phi}^{\prime 4}}{2M^6M_p^2a^4}, \nonumber \\
\tilde{\xi}&\coloneqq& \xi+\frac{\bar{\phi}^{\prime 2}}{2M^3M_pa^2}.
\end{eqnarray}

It is now straightforward to solve the equation of motion \eqref{master_equation}. Integrating once over $r$ and solving the quadratic equation for $\phi_{l,r}$ gives:\footnote{We drop the second branch of the solution, where the sign in front of the square root is reversed, because it doesn't converge for $r\rightarrow \infty$.}
\begin{equation}
\phi_{l,r} = -\frac{\lambda(\tau)a^2M^3}{4}r\left( 1 - \sqrt{1+\frac{r_V^3}{r^3}\frac{M(r, \tau)}{M_0}} \right), \label{solution}
\end{equation}
where we defined the total mass of the object $M_0$ and the mass inclosed in the radius $r$:
\begin{equation}
M(r, \tau)\coloneqq 4\pi\int_0^r\mathrm{d}r'\,r^{\prime 2}\rho_l(r', \tau).
\end{equation}
We also introduced the Vainsthein radius:
\begin{equation}
r_V(\tau) \coloneqq \left( \frac{2\tilde{\xi} M_0}{\pi M_p M^3\lambda^2(\tau)} \right)^{1/3}. \label{Vainshtein_radius}
\end{equation}
On large scales ($r\gg r_V$), the field gradient $\phi_{l,r}$ has a $1/r^2$ dependence , i.e.\ is unscreened. We use this unscreened solution in the Einstein equation \eqref{Poisson} and obtain:
\begin{equation}
\frac{\partial}{\partial r}\left(r^2 \phi_{N,r} \right)=\frac{a^2r^2}{2M_p^2}\rho_l(r)\alpha, \quad \text{where} \quad \alpha\coloneqq 1+\frac{\bar{\phi}^{\prime 2}\tilde{\xi}}{M^3a^2\lambda(\tau)M_p}.
\end{equation}
Therefore, the acceleration of a test particle due to the fifth force, $\vec{a}_5$, compared to the acceleration due to gravity, $\vec{a}_N$, becomes for $r\gg r_V$:
\begin{equation}
\frac{|\vec{a}_5|}{|\vec{a}_N|} =\frac{2\xi\tilde{\xi}}{\lambda(\tau)\alpha}.
\end{equation}
Observations of the linear growth of structure in our Universe make it unlikely that this fraction could be significantly larger than $1$, see Section \ref{sec:assumption1}, and also imply $\alpha= \mathcal{O}(1)$. Therefore, we assume in the following:
\begin{equation}
\frac{2\xi\tilde{\xi}}{\lambda(\tau)} \lesssim 1. \label{linear_force}
\end{equation}

In some situations the solution in Eq.\ \eqref{solution} can be analytically integrated once more in $r$. For example outside of the boundaries of the overdensity ($r>r_0$) the solution is given in terms of the Hypergeometric function $\prescript{}{2}F_1$:
\begin{equation}
\phi_l(r, \tau)=-\frac{\lambda(\tau)a^2M^3}{4}r^2\left( \frac{1}{2}-2\sqrt{1+\left( \frac{r_V(\tau)}{r} \right)^3} +\frac{3}{2}\prescript{}{2}F_1\left[ -\frac{2}{3}; \frac{1}{2}; \frac{1}{3}; -\left( \frac{r_V(\tau)}{r} \right)^3 \right] \right).
\end{equation}
Deep inside the Vainsthein radius ($r\ll r_V$) this solution is well approximated by:
\begin{equation}
\phi_l(r, \tau) \approx -\frac{\gamma}{2}\lambda(\tau)a^2M^3r_V^2(\tau), \label{inside_Vainshtein}
\end{equation}
where we abbreviated:
\begin{equation}
\gamma\coloneqq \frac{3\Gamma\left(\frac{1}{3}\right)\Gamma\left(\frac{7}{6}\right)}{4\sqrt{\pi}} \approx 1.0516\dots
\end{equation}
If the density profile of the overdensity is well described by a power law $\rho_l\propto r^{-\beta}$, the solution inside the overdensity becomes (for $\beta\neq 4$):
\begin{eqnarray}
\phi_l(r<r_0, \tau)= \phi_l(r_0, \tau) + \frac{\lambda(\tau)a^2M^3r_V^2}{4(2-\beta/2)}\left( \frac{r^{2-\beta/2}}{\sqrt{r_V}r_0^{3/2-\beta/2}} -\sqrt{\frac{r_0}{r_V}} \right).
\end{eqnarray}
If $\beta<4$ and the Vainshtein radius is much larger than the overdensity itself ($r_0\ll r_V$), i.e.\ the overdensity is screened, this solution is actually well approximated by just $\phi_l(r_0, \tau)$, which is given by Eq.\ \eqref{inside_Vainshtein}. $\beta<4$ is a very reasonable assumption for overdensities like the Milky Way which are typically assumed to have a NFW profile with $\beta=1$ in the center of the galaxy and $\beta=3$ on the outskirts of the galaxy.

We summarize, if $r_0\ll r_V$ and $\beta<4$, the solution for $\phi_l$ deep inside the Vainshtein radius is approximately given by:
\begin{equation}
\phi_l(r, \tau)\approx -\frac{\gamma}{2}\lambda(\tau)a^2M^3r_V^2(\tau)=8\gamma\frac{\tilde{\xi} M_p}{\lambda(\tau)}\phi_N(r_V),
\end{equation}
where the gravitational potential at $r_V$  is $\phi_N(r_V)=-M_0/8\pi M_p^2 r_V$. Taking $\log\Omega^{1/2}=\xi\phi/M_p$, we arrive at the analogue to Eq. \eqref{solar_system_estimate}:
\begin{equation}
\frac{1}{2}\log\frac{\Omega(r)}{\Omega_0}=\frac{1}{2}\log\left( 1+\frac{\omega(r)}{\Omega_0} \right) \approx 8\gamma\frac{\xi\tilde{\xi}}{\lambda(\tau)}\phi_N(r_V). \label{analogue}
\end{equation}
The bound in Eq.\ \eqref{linear_force} together with $|\phi_N(r_V)|\ll1$ requires the absolute value of the right-hand side of Eq.\ \eqref{analogue} to be small compared to $1$. Therefore, we can Taylor expand the left-hand side to find:
\begin{equation}
\omega(r)\approx 16\gamma\frac{\xi\tilde{\xi}}{\lambda(\tau)}\phi_N(r_V)\Omega_0. \label{galileon_estimate}
\end{equation}
We conclude that the relative deviation between the local and the cosmological conformal factor is in this case proportional to the gravitational potential at the Vainshtein radius $\phi_N(r_V)$, i.e.\ far outside the object. This is a far tighter constraint than the conservative result in Eq.\ \eqref{frame_equality}, where $\omega(r)$ is proportional to $\phi_N(r)$, the gravitational potential inside the object.

\bibliographystyle{JHEP}
\bibliography{constraining_conformal_factors}

\providecommand{\href}[2]{#2}\begingroup\raggedright\begin{thebibliography}{10}

\bibitem{Will:2014kxa}
C.~M. Will, \emph{{The Confrontation between General Relativity and
  Experiment}}, \href{https://doi.org/10.12942/lrr-2014-4}{\emph{Living Rev.
  Rel.} {\bfseries 17} (2014) 4}
  [\href{https://arxiv.org/abs/1403.7377}{{\ttfamily 1403.7377}}].

\bibitem{Adelberger:2009zz}
E.~G. Adelberger, J.~H. Gundlach, B.~R. Heckel, S.~Hoedl and S.~Schlamminger,
  \emph{{Torsion balance experiments: A low-energy frontier of particle
  physics}}, \href{https://doi.org/10.1016/j.ppnp.2008.08.002}{\emph{Prog.
  Part. Nucl. Phys.} {\bfseries 62} (2009) 102}.

\bibitem{Pulsars}
R.~A. {Hulse} and J.~H. {Taylor}, \emph{{Discovery of a pulsar in a binary
  system.}}, \href{https://doi.org/10.1086/181708}{\emph{ApJ} {\bfseries 195}
  (1975) L51}.

\bibitem{Monitor:2017mdv}
{\scshape LIGO Scientific, Virgo, Fermi-GBM, INTEGRAL} collaboration,
  \emph{{Gravitational Waves and Gamma-rays from a Binary Neutron Star Merger:
  GW170817 and GRB 170817A}},
  \href{https://doi.org/10.3847/2041-8213/aa920c}{\emph{Astrophys. J.}
  {\bfseries 848} (2017) L13}
  [\href{https://arxiv.org/abs/1710.05834}{{\ttfamily 1710.05834}}].

\bibitem{Ferreira:2019xrr}
P.~G. Ferreira, \emph{{Cosmological Tests of Gravity}},
  \href{https://doi.org/10.1146/annurev-astro-091918-104423}{\emph{Ann. Rev.
  Astron. Astrophys.} {\bfseries 57} (2019) 335}
  [\href{https://arxiv.org/abs/1902.10503}{{\ttfamily 1902.10503}}].

\bibitem{Akrami:2018vks}
{\scshape Planck} collaboration, \emph{{Planck 2018 results. I. Overview and
  the cosmological legacy of Planck}},
  \href{https://arxiv.org/abs/1807.06205}{{\ttfamily 1807.06205}}.

\bibitem{Verde:2019ivm}
L.~Verde, T.~Treu and A.~G. Riess, \emph{{Tensions between the Early and the
  Late Universe}},  in \emph{{Nature Astronomy 2019}}, 2019,
  \href{https://arxiv.org/abs/1907.10625}{{\ttfamily 1907.10625}},
  \href{https://doi.org/10.1038/s41550-019-0902-0}{DOI}.

\bibitem{Abbott:2017wau}
{\scshape DES} collaboration, \emph{{Dark Energy Survey year 1 results:
  Cosmological constraints from galaxy clustering and weak lensing}},
  \href{https://doi.org/10.1103/PhysRevD.98.043526}{\emph{Phys. Rev.}
  {\bfseries D98} (2018) 043526}
  [\href{https://arxiv.org/abs/1708.01530}{{\ttfamily 1708.01530}}].

\bibitem{Clifton:2011jh}
T.~Clifton, P.~G. Ferreira, A.~Padilla and C.~Skordis, \emph{{Modified Gravity
  and Cosmology}},
  \href{https://doi.org/10.1016/j.physrep.2012.01.001}{\emph{Phys. Rept.}
  {\bfseries 513} (2012) 1} [\href{https://arxiv.org/abs/1106.2476}{{\ttfamily
  1106.2476}}].

\bibitem{Joyce:2014kja}
A.~Joyce, B.~Jain, J.~Khoury and M.~Trodden, \emph{{Beyond the Cosmological
  Standard Model}},
  \href{https://doi.org/10.1016/j.physrep.2014.12.002}{\emph{Phys. Rept.}
  {\bfseries 568} (2015) 1} [\href{https://arxiv.org/abs/1407.0059}{{\ttfamily
  1407.0059}}].

\bibitem{Ishak:2018his}
M.~Ishak, \emph{{Testing General Relativity in Cosmology}},
  \href{https://doi.org/10.1007/s41114-018-0017-4}{\emph{Living Rev. Rel.}
  {\bfseries 22} (2019) 1} [\href{https://arxiv.org/abs/1806.10122}{{\ttfamily
  1806.10122}}].

\bibitem{Khoury:2003rn}
J.~Khoury and A.~Weltman, \emph{{Chameleon cosmology}},
  \href{https://doi.org/10.1103/PhysRevD.69.044026}{\emph{Phys. Rev.}
  {\bfseries D69} (2004) 044026}
  [\href{https://arxiv.org/abs/astro-ph/0309411}{{\ttfamily
  astro-ph/0309411}}].

\bibitem{Hinterbichler:2010es}
K.~Hinterbichler and J.~Khoury, \emph{{Symmetron Fields: Screening Long-Range
  Forces Through Local Symmetry Restoration}},
  \href{https://doi.org/10.1103/PhysRevLett.104.231301}{\emph{Phys. Rev. Lett.}
  {\bfseries 104} (2010) 231301}
  [\href{https://arxiv.org/abs/1001.4525}{{\ttfamily 1001.4525}}].

\bibitem{Vainshtein1972393}
A.~Vainshtein, \emph{To the problem of nonvanishing gravitation mass},
  \href{https://doi.org/https://doi.org/10.1016/0370-2693(72)90147-5}{\emph{Physics
  Letters B} {\bfseries 39} (1972) 393 }.

\bibitem{Babichev:2009ee}
E.~Babichev, C.~Deffayet and R.~Ziour, \emph{{k-Mouflage gravity}},
  \href{https://doi.org/10.1142/S0218271809016107}{\emph{Int. J. Mod. Phys.}
  {\bfseries D18} (2009) 2147}
  [\href{https://arxiv.org/abs/0905.2943}{{\ttfamily 0905.2943}}].

\bibitem{Bellini:2014fua}
E.~Bellini and I.~Sawicki, \emph{{Maximal freedom at minimum cost: linear
  large-scale structure in general modifications of gravity}},
  \href{https://doi.org/10.1088/1475-7516/2014/07/050}{\emph{JCAP} {\bfseries
  1407} (2014) 050} [\href{https://arxiv.org/abs/1404.3713}{{\ttfamily
  1404.3713}}].

\bibitem{Horndeski:1974wa}
G.~W. Horndeski, \emph{{Second-order scalar-tensor field equations in a
  four-dimensional space}},
  \href{https://doi.org/10.1007/BF01807638}{\emph{Int. J. Theor. Phys.}
  {\bfseries 10} (1974) 363}.

\bibitem{Deffayet:2011gz}
C.~Deffayet, X.~Gao, D.~A. Steer and G.~Zahariade, \emph{{From k-essence to
  generalised Galileons}},
  \href{https://doi.org/10.1103/PhysRevD.84.064039}{\emph{Phys. Rev.}
  {\bfseries D84} (2011) 064039}
  [\href{https://arxiv.org/abs/1103.3260}{{\ttfamily 1103.3260}}].

\bibitem{Frusciante:2019xia}
N.~Frusciante and L.~Perenon, \emph{{Effective Field Theory of Dark Energy: a
  Review}}, \href{https://doi.org/10.1016/j.physrep.2020.02.004}{\emph{Phys.
  Rept.} {\bfseries 857} (2020) 1}
  [\href{https://arxiv.org/abs/1907.03150}{{\ttfamily 1907.03150}}].

\bibitem{Nicolis:2008in}
A.~Nicolis, R.~Rattazzi and E.~Trincherini, \emph{{The Galileon as a local
  modification of gravity}},
  \href{https://doi.org/10.1103/PhysRevD.79.064036}{\emph{Phys. Rev.}
  {\bfseries D79} (2009) 064036}
  [\href{https://arxiv.org/abs/0811.2197}{{\ttfamily 0811.2197}}].

\bibitem{Renk:2017rzu}
J.~Renk, M.~Zumalacárregui, F.~Montanari and A.~Barreira, \emph{{Galileon
  gravity in light of ISW, CMB, BAO and H$_0$ data}},
  \href{https://doi.org/10.1088/1475-7516/2017/10/020}{\emph{JCAP} {\bfseries
  1710} (2017) 020} [\href{https://arxiv.org/abs/1707.02263}{{\ttfamily
  1707.02263}}].

\bibitem{Baker:2017hug}
T.~Baker, E.~Bellini, P.~G. Ferreira, M.~Lagos, J.~Noller and I.~Sawicki,
  \emph{{Strong constraints on cosmological gravity from GW170817 and GRB
  170817A}}, \href{https://doi.org/10.1103/PhysRevLett.119.251301}{\emph{Phys.
  Rev. Lett.} {\bfseries 119} (2017) 251301}
  [\href{https://arxiv.org/abs/1710.06394}{{\ttfamily 1710.06394}}].

\bibitem{Creminelli:2017sry}
P.~Creminelli and F.~Vernizzi, \emph{{Dark Energy after GW170817 and
  GRB170817A}},
  \href{https://doi.org/10.1103/PhysRevLett.119.251302}{\emph{Phys. Rev. Lett.}
  {\bfseries 119} (2017) 251302}
  [\href{https://arxiv.org/abs/1710.05877}{{\ttfamily 1710.05877}}].

\bibitem{Sakstein:2017xjx}
J.~Sakstein and B.~Jain, \emph{{Implications of the Neutron Star Merger
  GW170817 for Cosmological Scalar-Tensor Theories}},
  \href{https://doi.org/10.1103/PhysRevLett.119.251303}{\emph{Phys. Rev. Lett.}
  {\bfseries 119} (2017) 251303}
  [\href{https://arxiv.org/abs/1710.05893}{{\ttfamily 1710.05893}}].

\bibitem{Ezquiaga:2017ekz}
J.~M. Ezquiaga and M.~Zumalacárregui, \emph{{Dark Energy After GW170817: Dead
  Ends and the Road Ahead}},
  \href{https://doi.org/10.1103/PhysRevLett.119.251304}{\emph{Phys. Rev. Lett.}
  {\bfseries 119} (2017) 251304}
  [\href{https://arxiv.org/abs/1710.05901}{{\ttfamily 1710.05901}}].

\bibitem{Creminelli:2019kjy}
P.~Creminelli, G.~Tambalo, F.~Vernizzi and V.~Yingcharoenrat,
  \emph{{Dark-Energy Instabilities induced by Gravitational Waves}},
  \href{https://arxiv.org/abs/1910.14035}{{\ttfamily 1910.14035}}.

\bibitem{Noller:2020afd}
J.~Noller, \emph{{Cosmological constraints on dark energy in light of
  gravitational wave bounds}},
  \href{https://arxiv.org/abs/2001.05469}{{\ttfamily 2001.05469}}.

\bibitem{khoury_nogo}
J.~Wang, L.~Hui and J.~Khoury, \emph{{No-Go Theorems for Generalized Chameleon
  Field Theories}},
  \href{https://doi.org/10.1103/PhysRevLett.109.241301}{\emph{Phys. Rev. Lett.}
  {\bfseries 109} (2012) 241301}
  [\href{https://arxiv.org/abs/1208.4612}{{\ttfamily 1208.4612}}].

\bibitem{lunarlaser_new}
F.~Hofmann and J.~Müller, \emph{{Relativistic tests with lunar laser
  ranging}}, \href{https://doi.org/10.1088/1361-6382/aa8f7a}{\emph{Class.
  Quant. Grav.} {\bfseries 35} (2018) 035015}.

\bibitem{Barrow:1997qh}
J.~D. Barrow, \emph{{Varying G and other constants}},
  \href{https://doi.org/10.1007/978-94-011-5046-0\_8}{\emph{NATO Sci. Ser. C}
  {\bfseries 511} (1998) 269}
  [\href{https://arxiv.org/abs/gr-qc/9711084}{{\ttfamily gr-qc/9711084}}].

\bibitem{Braglia:2020iik}
M.~Braglia, M.~Ballardini, W.~T. Emond, F.~Finelli, A.~E. Gumrukcuoglu,
  K.~Koyama et~al., \emph{{A larger value for $H_0$ by an evolving
  gravitational constant}},  \href{https://arxiv.org/abs/2004.11161}{{\ttfamily
  2004.11161}}.

\bibitem{Ballesteros:2020sik}
G.~Ballesteros, A.~Notari and F.~Rompineve, \emph{{The $H_0$ tension: $\Delta
  G_N$ vs. $\Delta N_{\rm eff}$}},
  \href{https://arxiv.org/abs/2004.05049}{{\ttfamily 2004.05049}}.

\bibitem{Alvey:2019ctk}
J.~Alvey, N.~Sabti, M.~Escudero and M.~Fairbairn, \emph{{Improved BBN
  Constraints on the Variation of the Gravitational Constant}},
  \href{https://doi.org/10.1140/epjc/s10052-020-7727-y}{\emph{Eur. Phys. J. C}
  {\bfseries 80} (2020) 148}
  [\href{https://arxiv.org/abs/1910.10730}{{\ttfamily 1910.10730}}].

\bibitem{Ooba:2017gyn}
J.~Ooba, K.~Ichiki, T.~Chiba and N.~Sugiyama, \emph{{Cosmological constraints
  on scalar--tensor gravity and the variation of the gravitational constant}},
  \href{https://doi.org/10.1093/ptep/ptx046}{\emph{PTEP} {\bfseries 2017}
  (2017) 043E03} [\href{https://arxiv.org/abs/1702.00742}{{\ttfamily
  1702.00742}}].

\bibitem{Babichev:2011iz}
E.~Babichev, C.~Deffayet and G.~Esposito-Farese, \emph{{Constraints on
  Shift-Symmetric Scalar-Tensor Theories with a Vainshtein Mechanism from
  Bounds on the Time Variation of G}},
  \href{https://doi.org/10.1103/PhysRevLett.107.251102}{\emph{Phys. Rev. Lett.}
  {\bfseries 107} (2011) 251102}
  [\href{https://arxiv.org/abs/1107.1569}{{\ttfamily 1107.1569}}].

\bibitem{Kimura:2011dc}
R.~Kimura, T.~Kobayashi and K.~Yamamoto, \emph{{Vainshtein screening in a
  cosmological background in the most general second-order scalar-tensor
  theory}}, \href{https://doi.org/10.1103/PhysRevD.85.024023}{\emph{Phys. Rev.}
  {\bfseries D85} (2012) 024023}
  [\href{https://arxiv.org/abs/1111.6749}{{\ttfamily 1111.6749}}].

\bibitem{Barreira:2013xea}
A.~Barreira, B.~Li, C.~M. Baugh and S.~Pascoli, \emph{{Spherical collapse in
  Galileon gravity: fifth force solutions, halo mass function and halo bias}},
  \href{https://doi.org/10.1088/1475-7516/2013/11/056}{\emph{JCAP} {\bfseries
  1311} (2013) 056} [\href{https://arxiv.org/abs/1308.3699}{{\ttfamily
  1308.3699}}].

\bibitem{Brax:2018zvh}
P.~Brax and P.~Valageas, \emph{{Nonscreening of the cosmological background in
  K-mouflage modified gravity}},
  \href{https://doi.org/10.1103/PhysRevD.98.083509}{\emph{Phys. Rev. D}
  {\bfseries 98} (2018) 083509}
  [\href{https://arxiv.org/abs/1806.09414}{{\ttfamily 1806.09414}}].

\bibitem{Ferreira_galaxy}
H.~Desmond, P.~G. Ferreira, G.~Lavaux and J.~Jasche, \emph{{Fifth force
  constraints from the separation of galaxy mass components}},
  \href{https://doi.org/10.1103/PhysRevD.98.064015}{\emph{Phys. Rev.}
  {\bfseries D98} (2018) 064015}
  [\href{https://arxiv.org/abs/1807.01482}{{\ttfamily 1807.01482}}].

\bibitem{Burrage:2014oza}
C.~Burrage, E.~J. Copeland and E.~A. Hinds, \emph{{Probing Dark Energy with
  Atom Interferometry}},
  \href{https://doi.org/10.1088/1475-7516/2015/03/042}{\emph{JCAP} {\bfseries
  1503} (2015) 042} [\href{https://arxiv.org/abs/1408.1409}{{\ttfamily
  1408.1409}}].

\bibitem{Burrage:2017qrf}
C.~Burrage and J.~Sakstein, \emph{{Tests of Chameleon Gravity}},
  \href{https://doi.org/10.1007/s41114-018-0011-x}{\emph{Living Rev.\ Rel.}
  {\bfseries 21} (2018) 1} [\href{https://arxiv.org/abs/1709.09071}{{\ttfamily
  1709.09071}}].

\bibitem{Tully:2014gfa}
R.~B. Tully, H.~Courtois, Y.~Hoffman and D.~Pomarède, \emph{{The Laniakea
  supercluster of galaxies}},
  \href{https://doi.org/10.1038/nature13674}{\emph{Nature} {\bfseries 513}
  (2014) 71} [\href{https://arxiv.org/abs/1409.0880}{{\ttfamily 1409.0880}}].

\bibitem{Amendola:2004qb}
L.~Amendola, \emph{{Phantom energy mediates a long-range repulsive force}},
  \href{https://doi.org/10.1103/PhysRevLett.93.181102}{\emph{Phys.\ Rev.\
  Lett.} {\bfseries 93} (2004) 181102}
  [\href{https://arxiv.org/abs/hep-th/0409224}{{\ttfamily hep-th/0409224}}].

\bibitem{Wittner:2020yfc}
M.~Wittner, G.~Laverda, O.~F. Piattella and L.~Amendola, \emph{{Transient weak
  gravity in scalar-tensor theories}},
  \href{https://arxiv.org/abs/2003.08950}{{\ttfamily 2003.08950}}.

\bibitem{Amendola:2017orw}
L.~Amendola, M.~Kunz, I.~D. Saltas and I.~Sawicki, \emph{{Fate of Large-Scale
  Structure in Modified Gravity After GW170817 and GRB170817A}},
  \href{https://doi.org/10.1103/PhysRevLett.120.131101}{\emph{Phys.\ Rev.\
  Lett.} {\bfseries 120} (2018) 131101}
  [\href{https://arxiv.org/abs/1711.04825}{{\ttfamily 1711.04825}}].

\bibitem{DES_year1}
{\scshape DES} collaboration, \emph{{Dark Energy Survey Year 1 Results:
  Constraints on Extended Cosmological Models from Galaxy Clustering and Weak
  Lensing}}, \href{https://doi.org/10.1103/PhysRevD.99.123505}{\emph{Phys.
  Rev.} {\bfseries D99} (2019) 123505}
  [\href{https://arxiv.org/abs/1810.02499}{{\ttfamily 1810.02499}}].

\bibitem{Mueller:2016kpu}
E.-M. Mueller, W.~Percival, E.~Linder, S.~Alam, G.-B. Zhao, A.~G. Sánchez
  et~al., \emph{{The clustering of galaxies in the completed SDSS-III Baryon
  Oscillation Spectroscopic Survey: constraining modified gravity}},
  \href{https://doi.org/10.1093/mnras/stx3232}{\emph{Mon. Not. Roy. Astron.
  Soc.} {\bfseries 475} (2018) 2122}
  [\href{https://arxiv.org/abs/1612.00812}{{\ttfamily 1612.00812}}].

\bibitem{Barreira:2016ovx}
A.~Barreira, A.~G. Sánchez and F.~Schmidt, \emph{{Validating estimates of the
  growth rate of structure with modified gravity simulations}},
  \href{https://doi.org/10.1103/PhysRevD.94.084022}{\emph{Phys. Rev.}
  {\bfseries D94} (2016) 084022}
  [\href{https://arxiv.org/abs/1605.03965}{{\ttfamily 1605.03965}}].

\bibitem{Bose:2017myh}
B.~Bose, K.~Koyama, W.~A. Hellwing, G.-B. Zhao and H.~A. Winther,
  \emph{{Theoretical accuracy in cosmological growth estimation}},
  \href{https://doi.org/10.1103/PhysRevD.96.023519}{\emph{Phys. Rev.}
  {\bfseries D96} (2017) 023519}
  [\href{https://arxiv.org/abs/1702.02348}{{\ttfamily 1702.02348}}].

\bibitem{Equivalence_principle}
L.~Hui, A.~Nicolis and C.~Stubbs, \emph{{Equivalence Principle Implications of
  Modified Gravity Models}},
  \href{https://doi.org/10.1103/PhysRevD.80.104002}{\emph{Phys. Rev.}
  {\bfseries D80} (2009) 104002}
  [\href{https://arxiv.org/abs/0905.2966}{{\ttfamily 0905.2966}}].

\bibitem{Falck:2014jwa}
B.~Falck, K.~Koyama, G.-b. Zhao and B.~Li, \emph{{The Vainshtein Mechanism in
  the Cosmic Web}},
  \href{https://doi.org/10.1088/1475-7516/2014/07/058}{\emph{JCAP} {\bfseries
  1407} (2014) 058} [\href{https://arxiv.org/abs/1404.2206}{{\ttfamily
  1404.2206}}].

\bibitem{Hiramatsu:2012xj}
T.~Hiramatsu, W.~Hu, K.~Koyama and F.~Schmidt, \emph{{Equivalence Principle
  Violation in Vainshtein Screened Two-Body Systems}},
  \href{https://doi.org/10.1103/PhysRevD.87.063525}{\emph{Phys. Rev.}
  {\bfseries D87} (2013) 063525}
  [\href{https://arxiv.org/abs/1209.3364}{{\ttfamily 1209.3364}}].

\bibitem{Perenon:2019dpc}
L.~Perenon, J.~Bel, R.~Maartens and A.~de~la Cruz-Dombriz, \emph{{Optimising
  growth of structure constraints on modified gravity}},
  \href{https://doi.org/10.1088/1475-7516/2019/06/020}{\emph{JCAP} {\bfseries
  06} (2019) 020} [\href{https://arxiv.org/abs/1901.11063}{{\ttfamily
  1901.11063}}].

\bibitem{Wolf:2019hun}
W.~J. Wolf and M.~Lagos, \emph{{Standard Sirens as a Novel Probe of Dark
  Energy}}, \href{https://doi.org/10.1103/PhysRevLett.124.061101}{\emph{Phys.
  Rev. Lett.} {\bfseries 124} (2020) 061101}
  [\href{https://arxiv.org/abs/1910.10580}{{\ttfamily 1910.10580}}].

\bibitem{Lagos:2020mzy}
M.~Lagos and H.~Zhu, \emph{{Gravitational couplings in Chameleon models}},
  \href{https://arxiv.org/abs/2003.01038}{{\ttfamily 2003.01038}}.

\bibitem{Deffayet:2010qz}
C.~Deffayet, O.~Pujolas, I.~Sawicki and A.~Vikman, \emph{{Imperfect Dark Energy
  from Kinetic Gravity Braiding}},
  \href{https://doi.org/10.1088/1475-7516/2010/10/026}{\emph{JCAP} {\bfseries
  1010} (2010) 026} [\href{https://arxiv.org/abs/1008.0048}{{\ttfamily
  1008.0048}}].

\end{thebibliography}\endgroup
\end{document}